\documentclass[10pt,orivec]{llncs}

\usepackage{amsmath}
\usepackage{amssymb}
\usepackage{graphicx}
\usepackage{bbm}
\usepackage{stmaryrd}
\usepackage{epsfig}
\usepackage[final]{listings}
\usepackage{url}
\usepackage{xcolor}
\usepackage{enumerate}
\usepackage{txfonts}

\definecolor{mygreen}{rgb}{0,0.6,0}
\definecolor{myblue}{rgb}{0,0,0.6}
\lstdefinelanguage{ABS}{keywords={
    local, view, call, specifies,
    assert,this,dyndelta,new,VM,data,type,def,case,of,cog, class,interface,extends,implements,if,then,else,await,get, Fut,return,skip,while,module, import, export, from, suspend, 
    delta,adds,modifies,removes,original,productline,features,
    core,corefeatures,optionalfeatures,after,when,product,hasAttribute,
    hasMethod,root,extension,group,allof,oneof,require,exclude,original,
    ifin,ifout,opt,null,critical,port,rebind,duration,deadline,now,
    global,  view, result, local, specifies, grammar, send, receive},%
  sensitive=true, comment=[l]{//},
  morecomment=[s]{/*}{*/},
  morestring=[b]"}

\lstdefinestyle{absstyle}{
language=ABS,columns=fullflexible,
 		   mathescape=true,%
 		   showstringspaces=false,%
keywordstyle=\tt\color{myblue}, 
commentstyle=\tt\color{mygreen},%
basicstyle=\footnotesize\tttfamily,
inputencoding=latin1, 
extendedchars,xleftmargin=2em
}

\lstnewenvironment{abs}[1][]{%
 \lstset{style=absstyle,#1}%
   \csname lst@SetFirstLabel\endcsname}
  {\csname lst@SaveFirstLabel\endcsname}

\newcommand{\absinline}[1]{\lstinline[language=ABS,columns=fullflexible,mathescape=true,keywordstyle=\tt,basicstyle=\normalsize\ttfamily]|#1|} 

\lstnewenvironment{absexample}{
~\\
\noindent 
\small{\textcolor{green!80!black}{\bfseries\sffamily{Example:}}}\vspace{-4pt}
\lstset{style=absstyle,
basicstyle=\ttfamily,
framerule=.5pt,
backgroundcolor=\color{green!3},
rulecolor=\color{green!50!black},
frame=tblr,
xleftmargin=4pt,
xrightmargin=4pt,
}}
{\vspace{\baselineskip}}

\lstnewenvironment{absexamplen}[1][]{
\lstset{style=absstyle,
basicstyle=\footnotesize\ttfamily,
framerule=.5pt,
belowskip=0pt,
backgroundcolor=\color{green!3},
rulecolor=\color{green!50!black},
frame=tblr,
captionpos=b,
xleftmargin=4pt,
xrightmargin=4pt,#1
}}
{\vspace{\baselineskip}}

\newcommand{\bigfract}[2]{\frac{^{\textstyle #1}}{_{\textstyle #2}}}
\newcommand{\rulenamex}[1]{\mbox{\scriptsize\sc(#1)}}

\def \mathrule #1#2#3{\begin{array}{l} 
                       {\rulenamex{#1}}
                       \\ \bigfract{#2}{#3}
                      \end{array}}

\def \mathax #1#2{\begin{array}{l} 
                   {\mbox{\scriptsize {{\sc (#1)}}} } 
                   \\ #2
                  \end{array}}

\newcommand{\subst}[2]{\{\raisebox{.5ex}{\footnotesize$#1$}  /
                       \raisebox{-.5ex}{\footnotesize$#2$} \}}

\newcommand{\hole}{\mbox{$[~]$}}

\renewcommand{\bar}[1]{\overline{#1}}

\newcommand{\new}{\mbox{\tt new}~}
\newcommand{\newloc}{\mbox{\tt new~local}~}

\newcommand{\this}{{\tt this}}
\newcommand{\destiny}{{\tt destiny}}

\newcommand{\get}{{\tt get}}

\newcommand{\async}{\mbox{\tt !}}

\newcommand{\fut}{\textit{fut}}

\newcommand{\cog}{{\it cog}}
\newcommand{\invoc}{{\it invoc}}

\newcommand{\btnull}{\mbox{{\tt 0}}}

\newcommand{\slam}[1]{\raisebox{-.2ex}{\mbox{\large {\tt [}}} #1 \raisebox{-.2ex}{\mbox{\large {\tt ]}}}}

\newcommand{\eqdef}{\stackrel{\it def}{=}}

\newcommand{\cntb}{\mathbbm{b}}

\newcommand{\concntc}{\mathbbm{k}}
\newcommand{\mcntc}{\mathbbm{C}}

\newcommand{\strip}[1]{\lfloor #1 \rfloor}

\renewcommand{\emptyset}{\varnothing}

\newcommand{\coreABS}{\mbox{{\tt tml}}}

\newcommand{\key}[1]{\mbox{\ttfamily\bfseries #1}}

\newcommand{\many}[1]{\overline{#1}}

\newcommand{\feval}[2]{[\![#1]\!]_{#2}}

\newcommand{\nt}[1]{\textit{#1}}

\newcommand{\sep}{\mid}

\newcommand{\ntyperule}[3]{ 
  \begin{array}{c} 
    \textsc{\scriptsize ({#1})} \\ 
    #2 \\ 
    \hline
    #3 
  \end{array} }
 
\newcommand{\nredrule}[2]{ 
  \begin{array}{c} 
    \textsc{\scriptsize ({#1})} \\ 
    #2  
  \end{array}}

\newcommand{\stable}[1]{{\it stable}_{\, #1}}
\newcommand{\strongstable}[1]{{\it strongstable}_{\, #1}}

\newcommand{\job}[1]{\key{job}(#1)}
\newcommand{\synch}[1]{\Phi(#1)}

\newcommand{\rn}[1]{\mbox{\textsc{#1}}}        

\newcommand{\lbr}{\lsttxt{\{}}
\newcommand{\rbr}{\lsttxt{\}}}
\newcommand{\wrap}[1]{\lbr\, {#1}\, \rbr}
\newcommand{\ssemi}{\mathrel{\mbox{{\tt ;}}}}
\newcommand{\semi}{\mathrel{\mbox{{\tt ;}}}}

\newcommand{\nif}{\lstcmd{if}}

\newcommand{\nelse}{\lstcmd{else}}

\newcommand{\nreturn}{\lstcmd{return}}

\newcommand{\etrue}{\lsttxt{true}}
\newcommand{\efalse}{\lsttxt{false}}

\newcommand{\set}[1]{\wrap{#1}}

\newcommand{\nget}{\lstcmd{get}}

\newcommand{\nthis}{\lstcmd{this}}
\newcommand{\nidle}{\lstcmd{idle}}

\newcommand{\ethis}{\nthis}

\newcommand{\eget}[1]{#1 \seqpoint \nget}

\newcommand{\eif}[3]{\nif\ #1\ \wrap{#2}\ \nelse\ \wrap{#3}}

\newcommand{\ereturn}[1]{\nreturn\ #1}

\newcommand{\nand}{\mathrel{\text{\tt and}}}
\newcommand{\nor}{\mathrel{\text{\tt or}}}
\newcommand{\se}{\mathit{se}}
\newcommand{\nse}{\mathit{nse}}

\newcommand{\frt}{\mathbbm{t}}
\newcommand{\frx}{\mathbbm{x}}

\newcommand{\cntcel}{\mathbbm{a}}

\newcommand{\mty}[2]{{#1}({#2})}

\newcommand{\cogty}[2]{#1[#2]}
\newcommand{\nucogty}[2]{\nu #1[#2]}
\newcommand{\const}{k}

\newcommand{\newel}[2]{\nu {#1}{:}~{#2}}
\newcommand{\rmel}[1]{{#1}^\checkmark}

\newcommand{\typepe}[3]{{#1}\typep {#2}\typed {#3}}
\newcommand{\typeee}[4]{{#1}\typep {#2}\typed {#3}, \; \slam{{#4}}}
\newcommand{\typest}[3]{{#1}\typep {#2}\typed {#3}}

\newcommand{\cinvoc}[3]{{[{#3}]^{#2}_{#1}}}

\newcommand{\Class}{\name{Class}}
\newcommand{\m}{\name{m}}

\newcommand{\unit}{{\mbox{\tt -\!\!-}}}

\newcommand{\cls}{\trianglerighteq}

\newcommand{\atranslate}{\text{\tt translate}}

\newcommand{\fcost}{\fun{cost}}

\usepackage[section]{placeins}

\usepackage{mlrules}
\setmlrules{size=\footnotesize}
\renewcommand{\eget}[1]{#1 . \nget}

\begin{document}

\title{Time complexity of concurrent programs\thanks{Supported by the EU projects FP7-610582
   \textsc{Envisage}: Engineering Virtualized Services
   (\texttt{http:/$\!$/www.envisage-project.eu}).}}
\subtitle{-- a technique based on behavioural types -- }
\author{Elena Giachino\inst{1} \and Einar Broch Johnsen\inst{2} \and Cosimo Laneve\inst{1} \and Ka I Pun\inst{2}}
\institute{\mbox{Dept.~of Computer Science and Engineering, University of Bologna -- 
           INRIA FOCUS}
           \and
           Dept.~of Informatics, University of Oslo}

\maketitle

\begin{abstract}
We study the problem of automatically computing the time complexity of concurrent
object-oriented programs. 
To determine this complexity we use intermediate 
abstract descriptions that record relevant information for the time analysis 
(cost of statements, creations of objects, and concurrent operations), called 
\emph{behavioural types}. Then, we define a translation function that takes behavioural 
types and makes the parallelism explicit into so-called \emph{cost equations},
which are fed to an automatic 
off-the-shelf solver for obtaining the time \mbox{complexity}. 
%
%
\end{abstract}

\section{Introduction}
\label{sec:introduction}
Computing the cost of a sequential algorithm has always been a primary
question for every programmer, who learns the basic techniques in the
first years of their computer science or engineering curriculum. This
cost is defined in terms of the input values to the algorithm and
over-approximates the number of the executed instructions. 
In turn, given an appropriate abstraction of the CPU speed of a runtime
system, one can obtain the expected computation time of the
algorithm.
%

The computational cost of algorithms is particularly relevant in
mainstream architectures, such as the cloud. In that context, a
service is a concurrent program that must comply with a so-called
\emph{service-level agreement} (SLA) regulating the cost in time and
assigning penalties for its infringement~\cite{buyya09fgcs}. The
service provider needs to make sure that the service is able to meet
the SLA, for example in terms of the end-user response time, by
deciding on a resource management policy and determining the
appropriate number of virtual machine instances (or containers) and
their parameter settings (e.g., their CPU speeds).  To help service
providers make correct decisions about the resource management before
actually deploying the service, we need static analysis methods for
resource-aware services~\cite{HaehnleJohnsen15}. In previous work by
the authors, cloud deployments expressed in the formal modeling
language ABS~\cite{johnsen15jlamp} have used a combination of cost
analysis and simulations to analyse resource
management~\cite{ABHJSTW13}, and a Hoare-style proof system to reason
about end-user deadlines has been developed for sequential
executions~\cite{JohnsenPSTC15}.  In contrast, we are here interested
in statically estimating the computation time of concurrent services
deployed on the cloud with a given dynamic resource management policy.

Technically, this paper proposes a behavioural type system expressing the
resource costs associated with computations and study how these types
can be used to soundly calculate the time complexity of parallel
programs deployed on the cloud. 
To succinctly formulate this problem,
our work is developed for {\coreABS}, a small formally defined
concurrent object-oriented language which uses asynchronous communications
to trigger parallel activities. The language defines
virtual machine instances in terms of dynamically created concurrent
object groups with bounds on the number of cycles they can perform
per time interval. As we are
interested in the concurrent aspects of these computations, we
abstract from sequential analysis in terms of a statement
$\key{job}(e)$, which defines the number of processing cycles 
required by the instruction -- this is similar to the \texttt{sleep(n)} operation 
in {\tt Java}.

The analysis of behavioural types is defined by translating
them in a code that is adequate for an off-the-shelf solver --
the {\tt CoFloCo} solver~\cite{FMH14}. As a consequence, we are able to 
determine the computational cost of algorithms in a parametric way with respect to their
inputs.






%
%
%
%

\emph{Paper overview.}  The language is defined in
Section~\ref{sec:syntax} and we discuss restrictions that ease the
development of our technique in
Section~\ref{sec:restrictions}. Section~\ref{sec:typesystem} presents
the behavioural type system and Section~\ref{sec.analysis} explains the
analysis of computation time based on these behavioural types.  In
Section~\ref{sec:proofoutline} we outline our correctness proof of the
type system with respect to the cost equations.
In Section~\ref{sec:relatedwork} we discuss the relevant related 
work and in Section~\ref{sec:conclusions} we deliver concluding
remarks.


\section{The language {\coreABS}}
\label{sec:syntax}

The syntax and the semantics of
{\coreABS} are defined in the following two subsections; the
third subsection discusses a few examples. 

\paragraph{Syntax.}
A {\coreABS} program is a sequence of method definitions 
$T\ \name m(\vect{T\ x })\wrap{\vect{F\ y\ssemi}\ s}$, ranged over by $M$,
 plus a main body
$\wrap{ \vect{F \, z \, \ssemi}\  s' }\ \key{with}\ \const$.
In {\coreABS} we distinguish between \emph{simple types} $T$
which are either integers \name{Int} or classes \name{Class} (there is just one 
class in {\coreABS}), and
\emph{types} $F$, which also include \emph{future types}
\name{Fut<$T$>}. 
These future types let asynchronous method invocations be typed
(see below).
The notation $\vect{T \ x}$ denotes any finite sequence of \emph{variable declarations}
 $T \ x$.
The elements of the sequence are separated by commas.
When we write $\vect{T \ x \ssemi}$ we mean
a sequence $T_1\ x_1 \semi \cdots \semi$ $T_n\ x_n \semi$ when the sequence is not
empty; we mean the possibly empty sequence otherwise.

The syntax of statements $s$, expressions with side-effects $z$ and expressions
$e$ of {\coreABS} is defined by the following grammar:
\begin{syntax}
\simpleentry s []	
					x=z \bnfor \eif ess \bnfor \job{e} \bnfor \ereturn e 
					 \bnfor s\semi s  \quad [~]
\\
\simpleentry z [] 	e \bnfor e \async \name m(\vect e) \bnfor e.\name m(\vect x) 
					\bnfor \eget e  \bnfor
					\new{} {\tt Class} \ \key{with}\ e \bnfor \newloc{} {\tt Class}  \qquad 			[~]
\\
\simpleentry e []  \ethis 
\bnfor \se \bnfor \nse 															[~]
\end{syntax}%
A statement $s$ may be either one of the standard operations of an imperative language 
or the job statement $\job{e}$ that delays the continuation by
$e$ cycles of the machine executing it. 

An expression $z$ may change the state of the system. In particular,
it may be an \emph{asynchronous} method invocation of the form
$e \async \name m(\vect e)$, which does not suspend the caller's
execution.  When the value computed by the invocation is needed, the
caller performs a \emph{non-blocking} $\name{get}$ operation: if the
value needed by a process is not available, then an awaiting process
is scheduled and executed, i.e., \emph{await-get}.  Expressions $z$
also include standard synchronous invocations $e . \name m(\vect e)$
and $\newloc{} {\tt Class}$, which creates a new object. The intended
meaning is to create the object in the same machine -- called
\emph{cog} or \emph{concurrent object group} -- of the caller, thus
sharing the processor of the caller: operations in the same virtual
machine interleave their evaluation (even if in the following
operational semantics the parallelism is not explicit).
Alternatively, one can create an object on a different cog with   
$\new{} {\tt Class} \ \key{with}\ e$ thus letting methods execute in parallel. In this 
case, $e$ represents the capacity of the new cog, that is, the number of cycles the cog
can perform per time interval. We assume the presence of a special identifier \texttt{this.capacity}
that returns the capacity of the corresponding cog. 

A {\em pure} expression $e$ can be the reserved identifier
\absinline{this}
or an integer expression. 
Since the analysis in Section~\ref{sec.analysis} cannot deal with generic integer expressions,
we parse  expressions in a careful way. In particular we split them  into \emph{size expressions}
$\se$, which are expressions in Presburger arithmetics (this is a decidable fragment of
Peano arithmetics that only contains addition), and \emph{non-size expressions}
$\nse$, which are the other type of expressions. The syntax of size and non-size expressions
is the following:
\begin{syntax}
 \simpleentry \nse [] \const \bnfor x \bnfor \nse \leq \nse \bnfor \nse \nand \nse \bnfor \nse \nor \nse 
 \quad [~]
 \oris  \nse + \nse \bnfor \nse - \nse 
 \bnfor \nse \times \nse \bnfor \nse / \nse [~]\\
 \simpleentry \se [] ve \bnfor ve \leq ve \bnfor \se \nand \se \bnfor \se \nor \se [~]\\
 \simpleentry ve [] \const \bnfor x \bnfor ve + ve \bnfor \const \times ve [~]\\
\simpleentry \const []  \textit{rational  constants}   [~]
\end{syntax}%
In the paper, we assume that sequences of declarations $\vect{T \ x}$
and method declarations $\vect{M}$ do not contain duplicate names.  We
also assume that $\name{return}$ statements have no continuation.

\paragraph{Semantics.}
\label{sec:semantics}

The semantics of {\coreABS} is defined by a transition system whose states 
are \emph{configurations} \nt{cn} that are defined by the following 
syntax.  
{\footnotesize
\[
\begin{array}{rcl@{\qquad\qquad}rcl}
\nt{cn} & ::= & \varepsilon \sep \nt{fut}(f,\nt{val}) \sep  \nt{ob}(o,c,p,q) 
                \sep \nt{invoc}(o,f,\m,\many{v})    
&
\nt{act}&::=& o \sep
\varepsilon
\\
        & \sep &  \nt{cog}(c,\nt{act},\const)\sep \nt{cn}~\nt{cn}
	& \nt{val}&::=& \nt{v}\sep \bot
\\
\nt{p} & ::= & \wrap{l \mid s} \sep \key{idle} 
&l & ::= & [ \cdots , x\mapsto v , \cdots ] 
\\
q & ::= & \emptyset \sep \wrap{l \mid s} 
                \sep q~q 
& \nt{v} & ::= & o \sep f \sep \const
\end{array}
\]
}%

A \emph{configuration cn} is a set of concurrent object groups (cogs),
objects, invocation messages and futures, and the empty configuration
is written as $\varepsilon$.  The associative and commutative union
operator on configurations is denoted by whitespace.  A $\nt{cog}$ is
given as a term $\nt{cog}(c, \nt{act}, \const)$ where $c$ and $\const$
are respectively the identifier and the capacity of the cog, and
$\nt{act}$ specifies the currently active object in the cog.  An object is
written as $\nt{ob}(o, c, p, q)$, where $o$ is the identifier of the
object,~$c$ the identifier of the cog the object belongs to,~$p$ an
\emph{active process}, and~$q$ a pool of \emph{suspended processes}.
A \emph{process} is written as $\wrap{l \mid s}$, where $l$ denotes
local variable bindings and $s$ a list of statements.  An
\emph{invocation message} is a term $\nt{invoc}(o, f, \m, \many{v})$
consisting of the callee $o$, the future $f$ 
to which the result of the call is returned,  the method name $m$, and
the set of actual parameter values for the call.  A \emph{future}
$\nt{fut}(f, \nt{val})$ contains an identifier $f$ and a reply value
$\nt{val}$, where~$\bot$ indicates the reply value of the future has not been
received.

\begin{figure}[t]
{\scriptsize
  \hspace*{-2cm}
$$
\begin{array}{c}
  \qquad
  \ntyperule{Assign-Local}{
    x \in \dom(l)  \quad  v=\feval{e}{l}
  }{
    ob(o,c,\wrap{l \mid x=e\semi s},q) \\
    \to {ob(o,c,\wrap{l~[x\mapsto v] \mid s},q)}
  }

  \qquad

  \ntyperule{Cond-True}{	
    \etrue = \feval{e}{l} 
  }{ 
    ob(o,c,\wrap{l \mid \eif{e}{s_1}{s_2}\semi s},q)\\
    \to
    ob(o,c,\wrap{l \mid s_1\semi s},q)
  }
  
  \qquad

  \ntyperule{Cond-False}{
    \efalse = \feval{e}{l} 	
  }{
  ob(o,c,\wrap{l \mid  \eif{e}{s_1}{s_2}\semi s },q)
  \\
  \to
  ob(o,c,\wrap{l \mid s_2\semi s},q)
  }

  \\
  \\

  \ntyperule{New}{
    c' = \text{fresh}(\,) \quad o' = \text{fresh}(\,) \quad  
    \const=\feval{e}{l}
  }{
    ob(o,c,\wrap{l \mid x=\new{} \Class\ \key{with}\ e\semi s},q)
    \\
    \to ob(o,c,\wrap{l \mid x=o'\semi s},q)\ \
    ob(o',c',\nidle,\emptyset)\ \ cog(c', o',\const)
  }

  \qquad

  \ntyperule{New-Local}{
    o' = \text{fresh}(\,)
  }{
    ob(o,c,\wrap{l \mid x=\newloc{} \Class\semi s},q)\\
    \to 
    ob(o,c,\wrap{l \mid x=o'\semi s},q)\ \ ob(o',c,\nidle,\emptyset)
  }

  \\
  \\
  \ntyperule{Get-True}{
    f = \feval{e}{l} \quad v \not = \bot
  }{
    ob(o,c,\wrap{l \mid x=\eget e\semi s },q)\ \
    \fut(f, v)
    \\
    \to 
    ob(o,c,\wrap{l\mid x = v \semi s },q)\ \
  \fut(f, v) 
  }

  \qquad

  \ntyperule{Get-False}{f = \feval{e}{l} 
  }{
  ob(o,c,\wrap{l \mid x=\eget e\semi s },q)\ \ \fut(f, \bot)
  \\
  \to 
  ob(o,c,\nidle ,q \cup \wrap{l \mid x=\eget e\semi s })\ \ \fut(f, \bot)
  }

  \\
  \\

  \ntyperule{Self-Sync-Call}{
  o=\feval{e}{l} \quad 
  \many{v}=\feval{\many{e}}{l} \quad
  f' = l(\destiny)
  \\
  f= \text{fresh}(\,)\quad \wrap{l' \mid s'} = \text{bind}(o, f, \m, \many{v})
  }{
  ob(o,c,\wrap{l \mid x=e . \m(\many{e})\semi s},q)\ \\
  \to ob(o,c,\wrap{l' \mid s'\semi\key{cont}(f')},q \cup \wrap{ l \mid
    x=\eget{f}\semi s})\ \ \fut(f, \bot )
  }

  \qquad

  \ntyperule{Self-Sync-Return-Sched}{
  f = l'(\text{destiny})
  }{
  ob(o,c,\wrap{l \mid \key{cont}(f)},q \cup \wrap{l' \mid s})\\
  \to ob(o,c,\wrap{l' \mid s},q)
  }

  \\
  \\

  \ntyperule{Cog-Sync-Call}{
  o'=\feval{e}{l} \quad 
  \many{v}=\feval{\many{e}}{l} \quad
  f' = l(\destiny)
  \\
  f= \text{fresh}(\,)\quad \wrap{l' \mid s'} = \text{bind}(o', f, \m, \many{v})
  }{
  ob(o,c,\wrap{l \mid x=e . \m(\many{e})\semi s},q)\ \\
  ob(o',c,\nidle,q')\ \ \nt{cog}(c, o, \const)\\
  \to ob(o,c,\nidle,q \cup \wrap{ l \mid x=\eget{f} \semi s})\  \ \fut(f, \bot )
  \\
  ob(o',c, \wrap{l' \mid s'\semi\key{cont}(f')}, q')\ \ \nt{cog}(c, o', \const)
  }

  \qquad

  \ntyperule{Cog-Sync-Return-Sched}{
  f = l'(\text{destiny}) 
  }{
  ob(o,c,\wrap{l \mid \key{cont}(f)},q)\
  \nt{cog}(c, o, \const)\\
  ob(o', c, \nidle, q' \cup \wrap{l' \mid s'})\\
  \to ob(o,c,\nidle,q)\ \ \nt{cog}(c, o', \const)
  \\
  ob(o', c, \wrap{l' \mid s'}, q')
  }

  \\
  \\

  \ntyperule{Async-Call}{
  o'=\feval{e}{l}
  \quad \many{v}=\feval{\many{e}}{l}
  \quad
  f = \text{fresh}(~)
  }
  {ob(o,c,\wrap{l \mid x=e \async \m(\many{e})\semi s},q)\ 
  \\
  \to  ob(o,c,\wrap{l \mid x=f\semi s},q)\ \ \invoc(o',f,\m,\many{v})\
  \ \fut(f,\bot)}

  \qquad

  \ntyperule{Bind-Mtd}{
  \wrap{ l \mid s } =\text{bind}(o,f,\m,\many{v})
  }{
  ob(o,c,p,q)\ \ \invoc(o,f,\m,\many{v})\\\to ob(o,c,p,q \cup \wrap{ l \mid s } )
  }

  \\
  \\

  \nredrule{Release-Cog}{
  ob(o,c,\nidle,q)\ \ \nt{cog}(c, o,\const)\\\to
  ob(o,c,\nidle, q)\ \ \nt{cog}(c, \varepsilon,\const)
  }
  \qquad
  \nredrule{Activate}{
  ob(o,c,\nidle,q \cup \wrap{ l \mid s})\ \ \nt{cog}(c, \varepsilon,\const) 
  \\ \to
  ob(o,c,\wrap{ l \mid s},q)\ \ \cog(c, o,\const) 
  }

  \\
  \\

  \ntyperule{Return}{
  v = \feval{e}{l}  \quad f = l(\text{destiny}) }
  {ob(o,c,\wrap{l \mid \key{return}\ e},q)\ \ \fut(f,\bot)\\
  \to ob(o,c,\nidle,q)\ \ \fut(f,v)
  }

  \qquad

  \ntyperule{Job-0}{\feval{e}{l}=0 }
  {ob(o,c,\wrap{l \mid \key{job}(e)\semi s},q) \\
  \to ob(o,c,\wrap{l \mid s},q) 
  }

  \qquad

  \ntyperule{Context}{
  \nt{cn} \to \nt{cn}'
  }{
  \nt{cn}  ~\nt{cn}''
  \to \nt{cn}' ~ \nt{cn}''
  }
\end{array}
$$
}
\caption{The transition relation of {\coreABS} -- part 1.}\label{fig:red-to}  
\end{figure}
%

The following auxiliary function is used in the semantic rules for invocations. 
Let $T' \ \m (\vect{T \ x}) \wrap{ \vect{F \ x'} ; s}$ 
be a method
declaration. Then 
\[
\text{bind}(o,f,\m,\many{v}) = \wrap{ 
[\text{destiny} \mapsto f, \bar{x} \mapsto \bar{v}, \bar{x'} \mapsto \bot] \mid s
\subst{o}{\this} }
\]

The \emph{transition rules} of {\coreABS} are given
in Figures~\ref{fig:red-to} and~\ref{fig:sem}.  
%
We discuss the most relevant ones: object creation, method invocation,
and the $\job{e}$ operator.
The creation of objects is handled by rules \rn{New} and
\rn{New-Local}: the former creates a new object inside a new cog with a given
capacity $e$, the latter creates an object
in the local cog.
Method invocations can be either synchronous or
asynchronous.  Rules \rn{Self-Sync-Call} and \rn{Cog-Sync-Call} 
specify synchronous invocations on objects belonging to the same cog of the 
caller. Asynchronous invocations can be performed on every object.

In our model, the unique operation that consumes time is $\job{e}$.
We notice that the reduction rules of Figure~\ref{fig:red-to} are not
defined for the $\job{e}$ statement, except the trivial case when the
value of $e$ is 0.  This means that time does not advance while
non-job statements are evaluated.  When the configuration $\nt{cn}$
reaches a \emph{stable} state, \emph{i.e.,}~no other transition is
possible apart from those evaluating the $\job{e}$ statements, then
the time is advanced by the minimum value that is necessary to let at
least \emph{one} process start.  In order to formalize this semantics, we
define the notion of stability and the \emph{update operation} of a
configuration $\nt{cn}$ (with respect to a time value $t$). Let
$\feval{e}{l}$ return the value of $e$ when variables are bound to
values stored in $l$.
\begin{definition}
\label{def:stability}
Let $t > 0$. A configuration $\nt{cn}$ is $t$-\emph{stable}, written
$\stable{t}(\nt{cn})$, if any object in $\nt{cn}$ is in one of the following
forms:
  \begin{enumerate}
    \item \label{def:stability.job}$ob(o, c, \wrap{l \mid {\tt job}(e) ; s}, q)$ with 
    $\nt{cog}(c, o, \const) \in \nt{cn}$ and $\feval{e}{l}/\const \geq t$,
    \item $ob(o, c, \nidle, q)$ and 
      \begin{itemize}\label{def:stability.idle}
        \item[i.] either $q = \emptyset$, 
        \item[ii.] or, for every $p \in
          q$, $p = \wrap{l \mid x=\eget{e} ; s}$ with $\feval{e}{l}=f$ and
          $\fut(f, \bot)$,
        \item[iii.] or, $\cog(c, o',\const) \in \nt{cn}$ where $o \neq
          o'$, and $o'$ satisfies Definition
            \ref{def:stability}.\ref{def:stability.job}.
      \end{itemize}
  \end{enumerate}
A configuration $\nt{cn}$ is \emph{strongly} $t$-\emph{stable}, written
$\strongstable{t}(\nt{cn})$, if it is $t$-stable and there is an object 
$ob(o, c, \wrap{l \mid {\tt job}(e) ; s}, q)$ with 
    $\nt{cog}(c, o, \const) \in \nt{cn}$ and $\feval{e}{l}/\const~=~t$.
\end{definition}
\begin{figure}[t]
{\footnotesize 
$
\begin{array}{c}
  \ntyperule{Tick}{
	\strongstable{t}(\nt{cn})
	}{
	\nt{cn} \to \synch{cn,t} 
  }
\\
\text{where} \hfill 
\\
\synch{cn,t} \; = \; \left\{
  \begin{array}{l@{\qquad}l}
        ob(o,c,\{l' \mid \job{\const'}\semi s\}, q)~\synch{cn',t}
        &
        \text{if}\; cn = ob(o,c,\{l \mid \job{e} \semi s\}, q) \ cn'
        \\
        & 
          \text{and} \; \cog(c,o,\const) \in \nt{cn}'
        \\
        & \text{and} \;
        \const' = \feval{e}{l}-{\const\mathop{*}t} 
\\
\\
		ob(o,c,\nidle,q)~\synch{cn',t}
        &
        \text{if}\; \nt{cn} = ob(o,c,\nidle,q)~\nt{cn}'
\\\\
        \nt{cn}
        &
        \text{otherwise.}
\end{array} \right.
\end{array}
$}
\caption{The transition relation of {\coreABS} -- part 2: the strongly stable case} \label{fig:sem} 
\end{figure}
Notice that  $t$-stable (and, consequently,  strongly $t$-stable) 
configurations cannot progress anymore because every object is stuck either on a job
or on unresolved get statements.
The update of $\nt{cn}$ with respect to a time value $t$, noted $\synch{cn,t}$ 
is defined in Figure~\ref{fig:sem}. Given these two notions, rule \rn{Tick} defines
the time progress. 


The initial configuration of a program with main method
$\wrap{\vect{F\ x ;}\ s} \ \key{with}\ \const
$ is  
\[
\begin{array}{l}
{\it ob}({\it start}, \text{start}, \wrap{ [ 
\text{destiny} \mapsto f_{\it start}, 
\bar{x} \mapsto \bot] \, | \, s }, \varnothing) \\ \cog(\text{start},{\it start},\const)
\end{array}
\]
 where
$\text{start}$ and ${\it start}$ are special cog and object names, respectively, 
and $f_{\it start}$ is a fresh future name.
As usual, $\to^*$ is the reflexive and transitive closure of $\to$. 



\paragraph{Examples.} ~\label{sec.fib.ex} To begin with, we discuss
the Fibonacci method. It is well known that the computational cost of
its sequential recursive implementation is exponential. However, this
is not the case for the parallel implementation. Consider
{\footnotesize
\begin{verbatim}
     Int fib(Int n) { 
            if (n<=1) { return 1; }
            else {  Fut<Int> f; Class z; Int m1; Int m2;
                    job(1);
                    z = new Class with this.capacity ;
                    f = this!fib(n-1); g = z!fib(n-2);
                    m1 = f.get; m2 = g.get;
                    return  m1 + m2;
            }
     }
\end{verbatim}  
}
%
%
%
\noindent
Here, the recursive invocation {\tt fib(n-1)} is performed on the \texttt{this} 
object while the invocation {\tt fib(n-2)} is performed
on a new cog with the same capacity (i.e., the object referenced by \texttt{z} is
created in a new cog set up with \texttt{this.capacity}),
which 
means that it can be performed in parallel with the former one. 
It turns out that the cost of the following invocation is {\tt n}. 
{\footnotesize
\begin{verbatim}
     Class z; Int m; Int x;
     x = 1;
     z = new Class with x;
     m = z.fib(n);
\end{verbatim}}
\noindent 
Observe that, by changing the line
{\tt x = 1;} into {\tt x =  2;} 
we obtain a cost of {\tt n/2}. 

Our semantics does not exclude paradoxical behaviours of programs 
that perform infinite actions 
without consuming time (preventing rule \rn{Tick} to apply), such as this one
{\footnotesize
\begin{verbatim}
     Int foo() { Int m; m = this.foo(); return m; }
\end{verbatim}  
}
\noindent
This kind of behaviours are well-known in the literature,  
(\emph{cf.}~Zeno behaviours) and they may be easily excluded 
from our analysis by constraining recursive invocations to be prefixed by a 
$\job{e}$-statement, with a positive $e$.
It is worth to observe that this condition is not sufficient to
eliminate paradoxical behaviours.
For instance the method below does not
terminate and, when invoked with {\tt this.fake(2)}, where {\tt this}
is in a cog of capacity 2, has cost 1.  
{\footnotesize
\begin{verbatim}
     Int fake(Int n) {
           Int m; Class x; 
           x = new Class with 2*n; job(1); m = x.fake(2*n); return m;
     }
\end{verbatim}  
}
\noindent
Imagine a parallel invocation of the following method
  {\footnotesize
\begin{verbatim}
     Int one() { job(1); }
\end{verbatim}}%
\noindent
on an object residing in a cog of capacity 1.
  At each stability point the $\job{1}$ of the latter method will
  compete with the $\job{1}$ of the former one, which will win every
  time, since having a greater (and growing) capacity it will require
  always less time. So at the first stability point we get
  $\job{1-1/2}$ (for the method \absinline{one}), then $\job{1-1/2-1/4}$ and so on, thus this sum will never reach 0.

In the examples above, the statement $\job{e}$ is a cost annotation that specifies
how many processing cycles are needed by the subsequent statement in
the code. We notice that this operation can also be used to program a timer which suspends
the current execution for $e$ units of time. For instance, let
{\footnotesize
\begin{verbatim}
     Int wait(Int n) { job(n); return 0; }
\end{verbatim}  
}
\noindent
Then, invoking {\tt wait} on an object with capacity 1
{\footnotesize
\begin{verbatim}
     Class timer; Fut<Class> f; Class x;
     timer = new Class with 1;
     f = timer!wait(5); x = f.get;
\end{verbatim}  
}
\noindent
one gets the suspension of the current thread for 5 units of time.

\section{Issues in computing the cost of {\coreABS} programs}
\label{sec:restrictions}

The computation time analysis of {\coreABS} programs is demanding. To highlight the
difficulties, we discuss a number of methods.
%
%
  {\footnotesize
\begin{verbatim}
    Int wrapper(Class x) {
          Fut<Int> f; Int z;
          job(1) ; f = x!server(); z = f.get;
          return z;    
    }
\end{verbatim}  
}
\noindent
Method \absinline{wrapper} performs an invocation on its argument \absinline{x}. In order
to determine the cost of \absinline{wrapper}, we notice that, if \absinline{x} is in 
the same cog of the carrier, then its cost is (assume that the capacity of the carrier is
{\tt 1}): ${\tt 1} + {\it cost}({\tt server})$ because the two invocations are
sequentialized. However, if the cogs of {\tt x} and of the carrier are different, then 
we are not able to compute the cost because we have no clue about the state of the cog
of {\tt x}.
Next consider the following definition of \absinline{wrapper}
  {\footnotesize
\begin{verbatim}
    Int wrapper_with_log(Class x) {
          Fut<Int> f; Fut<Int> g; Int z;
          job(1) ; f = x!server(); g = x!print_log(); z = f.get;
          return z;    
    }
\end{verbatim}  
}
In this case the wrapper also asks the server to print its log and this invocation is
not synchronized. We notice that the cost of \absinline{wrapper_with_log} is not 
anymore ${\tt 1} + {\it cost}({\tt server})$ (assuming that \absinline{x} is in the
same cog of the carrier) because \absinline{print_log} might be executed \emph{before}
{\tt server}. Therefore the cost of  \absinline{wrapper_with_log} is 
${\tt 1} + {\it cost}({\tt server}) + {\it cost}({\tt print\_log})$.

Finally, consider the following wrapper that 
also logs the information received from the server on a new cog without synchronising with it:
\\
\\
\begin{minipage}[h]{0.5\linewidth}
  {\footnotesize
\begin{verbatim}
    Int wrapper_with_external_log(Class x) {
          Fut<Int> f; Fut<Int> g; Int z; Class y;
          job(1) ; f = x!server(); g = x!print_log(); z = f.get;
          y = new Class with 1;
          f = y!external_log(z) ;
          return z;    
    }
\end{verbatim}  
}
\end{minipage}
\\
\\
What is the cost of \absinline{wrapper_with_external_log}? Well, the answer here is
debatable: one might discard the cost of  \absinline{y!external_log(z)} because it is useless for the
value returned by \absinline{wrapper_with_external_log}, or one might count it because
one wants to count every computation that has been triggered by a method in its 
cost. In this paper we adhere to the second alternative; however, we think that a 
better solution should be to return different cost for a method: a \emph{strict cost},
which spots the cost that is necessary for computing the returned value, and an
\emph{overall cost}, which is the one computed in this paper.

%
%
%

Anyway, by the foregoing discussion, 
as an initial step towards the time analysis of {\coreABS} programs, we simplify
our analysis by imposing the following constraint:
\begin{itemize}
\item[--]
\emph{it is possible to invoke methods on objects either in the same cog of the caller or 
on newly created cogs}. 


%
\end{itemize}
The above constraint means that, if the callee of an invocation is one
of the arguments of a method then it must be in the same cog of the
caller. It also means that, if an invocation is performed on a
returned object then this object must be in the same cog of the
carrier.  We will enforce these constraints in the typing system of
the following section -- see rule \rn{T-Invoke}.
%
%
%
%


 \section{A behavioural type system for {\coreABS}}
 \label{sec:typesystem}
 
In order to analyse the computation time of {\coreABS} programs we use abstract descriptions, 
called \emph{behavioural types}, which are intermediate codes highlighting the features of {\coreABS} programs that are relevant for the  analysis  
in Section~\ref{sec.analysis}. These abstract descriptions support compositional reasoning
and are associated to programs by means of a type system.
The syntax of behavioural types is defined as follows:
{\footnotesize
  \begin{syntax}
 \simpleentry 
    \frt [] \quad \unit \bnfor \se \bnfor \cogty{c}{\se}
    [basic value]   
    \\
    \simpleentry 
    \frx [] \quad f \bnfor \frt
    [extended value]
    \\
    \simpleentry
    \cntcel [] \quad 
    e \bnfor \nucogty{c}{\se} \bnfor
    \mty{\m}{\vect{\frt}} \rightarrow \frt \bnfor
    \newel{f}{\mty{\m}{\vect{\frt}} \rightarrow \frt} \bnfor 
    \rmel{f} \qquad 
    [atom]
    \\
    \simpleentry 
    \cntb [] \quad 
    \cntcel \triangleright \Gamma \bnfor \cntcel
    \fatsemi \cntb \bnfor (\se) \wrap{\cntb} \bnfor \cntb + \cntb 
    \qquad \qquad [behavioural type]
  \end{syntax}%
}%
where $c$, $c'$, $\cdots$ range over cog names and $f$, $f'$, $\cdots$
range over future names.  Basic values $\frt$ are either generic
(non-size) expressions $\unit$ or size expressions $\se$ or the
type $\cogty{c}{\se}$ of an object of cog $c$ with capacity
$\se$.  The extended values add future names to basic values.

Atoms $\cntcel$ define creation of cogs
($\nucogty{c}{\se}$), 
synchronous and asynchronous method invocations ($\mty{\m}{\vect{\frt}}\rightarrow\frt$ and
$\newel{f}{
  \mty{\m}{\vect{\frt}}\rightarrow\frt}$, respectively), and 
synchronizations on asynchronous invocations ($\rmel{f}$). We observe that cog creations
always carry a capacity, which has to be a size expression because our analysis in the
next section cannot deal with generic expressions. 
Behavioural types $\cntb$ are sequences of atoms
$\cntcel \fatsemi \cntb'$ or conditionals, typically
$(\se) \wrap{\cntb} + (\neg \se) \wrap{\cntb'}$ or $\cntb + \cntb'$,
according to whether the boolean guard is a size expression that
depends on the arguments of a method or not.  In order to type
sequential composition in a precise way (see rule~\rn{T-Seq}), the
leaves of behavioural types are labelled with \emph{environments},
ranged over by $\Gamma$, $\Gamma'$, $\cdots$.
Environments
are maps from method names {\m} to terms $(\vect{\frt}) \rightarrow \frt$, 
from variables to extended values $\frx$, and from future names to 
values that are either $\frt$ or $\rmel{\frt}$.

The abstract behaviour of methods is defined by \emph{method
  behavioural types} of the form: $\mty{\m}{\frt_t, \vect{\frt}}
\wrap{\cntb}: \frt_r$, where $\frt_t$ is the type value of the
receiver of the method, $\vect{\frt}$ are the type value of the
arguments, $\cntb$ is the abstract behaviour of the body, and $\frt_r$
is the type value of the returned object.  The subterm $\frt_t,
\vect{\frt}$ of the method contract is called \emph{header}; $\frt_r$
is called \emph{returned type value}. We assume that names in the
header occur linearly. Names in the header \emph{bind} the names in
$\cntb$ and in $\frt_r$.  The header and the returned type value,
written $(\frt_t,\vect{\frt}) \rightarrow \frt_r$, are called
\emph{behavioural type signature}. Names occurring in~$\cntb$
or~$\frt_r$ may be \emph{not bound} by header. These \emph{free names}
correspond to new cog creations and will be replaced by fresh cog
names during the analysis.  We use $\mcntc$ to range over method
behavioural types.

The type system uses judgments of the following form:
\begin{itemize}
\item[--] $\typepe{\Gamma}{e}{\frx}$
for pure expressions $e$, $\typepe{\Gamma}{f}{\frt}$ or $\typepe{\Gamma}{f}{\rmel{\frt}}$ 
for future names $f$, and  $\typepe{\Gamma}{\mty{\m}{\vect{\frt}}}{\frt}$
for methods.
\item[--]
$\typeee{\Gamma}{\nt{z}}{\frx}{\cntcel \triangleright \Gamma'}$
for expressions with side effects $\nt{z}$, where $\frx$ is the value, $\cntcel
\triangleright \Gamma'$ is the corresponding behavioural type,
where $\Gamma'$ is the environment $\Gamma$ \emph{with possible updates} 
of variables and future names. 
\item[--]
$\typest{\Gamma}{s}{\cntb}$, in this case the updated
environments  $\Gamma'$ are inside the
behavioural type, in correspondence of every branch of its. 
\end{itemize}

Since $\Gamma$ is a function, we use the standard predicates $x \in \dom(\Gamma)$ or 
$x \not \in \dom(\Gamma)$. Moreover, we define

\smallskip
$\begin{array}{@{\qquad}l@{\qquad}l}
\Gamma[x \mapsto \frx](y) \; \eqdef \;
	\left\{ \begin{array}{ll}
			 \frx & \text{if} \; y=x
			 \\ 
			 \Gamma(y)\ & \text{otherwise}
			 \end{array} \right.
\end{array}$
\smallskip

The \emph{multi-hole contexts} ${\cal C}[~]$ 
are defined
by the following syntax:

\smallskip
$\qquad{\cal C}\hole \; ::= \quad 
	\hole 
	\quad | \quad  \cntcel \fatsemi {\cal C}\hole
	\quad | \quad  {\cal C}\hole + {\cal C}\hole 
	\quad | \quad  (se) \wrap{{\cal C}\hole} 
        $
\smallskip

\noindent
and, whenever $\cntb = {\cal C}[\cntcel_1 \triangleright \Gamma_1]
\cdots [\cntcel_n \triangleright \Gamma_n]$, then $\cntb[x \mapsto
\frx]$ is defined as ${\cal C}[\cntcel_1 \triangleright \Gamma_1[x
\mapsto\frx]] \cdots [\cntcel_n \triangleright \Gamma_n[x
\mapsto \frx]]$.

The typing rules for expressions are defined in
Figure~\ref{fig:typesystem}. 
\begin{figure}[t]
{\footnotesize
\[
	\begin{array}{@{\hspace*{-.5cm}}c}
    \mathrule{T-Var}{
      x\in\dom(\Gamma)
    }{
      \typepe{\Gamma}{x}{\Gamma(x)}
    }
    \qquad
    \mathax{T-Se}{
      \typepe{\Gamma}{\se}{\se}
    }
    \qquad
    \mathax{T-Nse}{
      \typepe{\Gamma}{\nse}{\unit}
    }
    \qquad
    \mathrule{T-Method}{
      \begin{array}{c}
      \Gamma(\m) = (\vect{\frt}) \rightarrow \frt' 
      \\
      \fv(\frt') \setminus \fv(\vect{\frt}) \neq \emptyset
      \quad \text{implies} \quad \sigma(\frt') \fresh
      \end{array}
    }{
      \typepe{\Gamma}{
        \mty{\m}{\sigma(\vect{\frt})}
      }{\sigma(\frt')} 
    }
    \\\\
    \mathrule{T-New}{
      \typepe{\Gamma}{e}{\se}
      \quad
      c \fresh
    }{
      \typeee{\Gamma}{\new \Class\ \key{with}\ e}{\cogty{c}{\se}}{\nucogty{c}{\se}\triangleright 
        \Gamma[c\mapsto \se]} 
    }
    \qquad
    \mathrule{T-New-Local}{
      \typepe{\Gamma}{\ethis}{ \cogty{c}{\se} }
    }{
      \typeee{\Gamma}{\newloc \Class}{\cogty{c}{\se}}{\btnull\triangleright 
        \Gamma} 
    }
    \\\\
    \mathrule{T-Invoke-Sync}{
    	\begin{array}{c}
      \typepe{\Gamma}{e}{\cogty{c}{\se}}
      \quad
      \Gamma(\this)=\cogty{c}{\se}
      \\
      \typepe{\Gamma}{\vect{e}}{\vect{\frt}}
      \quad
      \typepe{\Gamma}{\mty{\m}{\cogty{c}{\se},\vect \frt}}{\frt'}
      \end{array}
    }{
      \typeee{\Gamma}{e.\name m(\vect e)}{\frt'}{\mty{\m}{\cogty{c}{\se},\vect{\frt}}\rightarrow \frt'\triangleright 
        \Gamma} 
    }
\qquad
    \mathrule{T-Invoke}{
    \begin{array}{c}
      \typepe{\Gamma}{e}{\cogty{c}{\se}}
      \quad
      (c \in \dom(\Gamma) \quad \mbox{or} \quad \Gamma(\ethis)=\cogty{c}{\se})
      \\
      \typepe{\Gamma}{\vect{e}}{\vect{\frt}}
      \quad
      \typepe{\Gamma}{\mty{\m}{\cogty{c}{\se},\vect{\frt}}}{\frt'}
      \quad
      f \fresh
      \end{array}
    }{
      \typeee{\Gamma}{e \async \name m(\vect e)}{f}{\newel{f}{\mty{\m}{\cogty{c}{\se},\vect{\frt}}\rightarrow \frt'} \triangleright 
        \Gamma[f \mapsto \frt']} 
    }
      \\\\
    \mathrule{T-Get}{
      \typepe{\Gamma}{e}{f}
      \quad
      \Gamma(f) = \frt    }{
      \typeee{\Gamma}{\eget e}{\frt}{\rmel{f}\triangleright 
        \Gamma[f\mapsto \rmel{\frt}]} 
    }
    \qquad
    \mathrule{T-Get-Top}{
      \typepe{\Gamma}{e}{f}
      \quad
      \Gamma(f) = \rmel{\frt}
    }{
      \typeee{\Gamma}{\eget e}{\frt}{\btnull \triangleright 
        \Gamma} 
    }
  \end{array}
  \]}
  \caption{Typing rules for expressions}
  \label{fig:typesystem}
\end{figure}
These rules are not standard because (size) expressions containing
method's arguments are typed with the expressions themselves. This is
crucial to the cost analysis in Section~\ref{sec.analysis}.  In
particular, \emph{cog creation} is typed by rule~\rn{T-New}, with
value $\cogty{c}{\se}$, where $c$ is the fresh name associated with
the new cog and $\se$ is the value associated with the declared
capacity.  The behavioural type for the cog creation is
$\nucogty{c}{\se}\triangleright \Gamma[c\mapsto{\se}]$, where the
newly created cog is added to $\Gamma$.  In this way, it is possible
to verify whether the receiver of a method invocation is within a
locally created cog or not by testing whether the receiver belongs to
$\dom(\Gamma)$ or not, respectively (\emph{cf.}~rule~\rn{T-Invoke}).
\emph{Object creation} (\emph{cf.}~rule~\rn{T-New-Local}) is typed as
the cog creation, with the exception that the cog name and the
capacity value are taken from the local cog and the behavioural type
is empty.  Rule \rn{T-Invoke} types \emph{method invocations}
$e!\m(\bar{e})$ by using a fresh future name $f$ that is associated to
the method name, the cog name of the callee and the arguments. In the
updated environment, $f$ is associated with the returned value.  Next
we discuss the constraints in the premise of the rule. As we discussed
in Section~\ref{sec:semantics}, asynchronous invocations are allowed
on callees located in the current cog, $\Gamma(\this)=\cogty{c}{\se}$,
or on a newly created object which resides in a fresh cog,
$c\in\dom(\Gamma)$.
%
%
%
Rule \rn{T-Get} defines the \emph{synchronization} with a method invocation that 
corresponds to a future $f$. The expression is typed with the value $\frt$ 
of $f$ in the environment and behavioural type $\rmel{f}$. $\Gamma$ is then updated for recording that
the synchronization has been already performed, thus any subsequent
synchronization on the same value would not imply any waiting time
(see that in rule \rn{T-Get-Top} the behavioural type is
$\btnull$).
The \emph{synchronous method invocation} in rule \rn{T-Invoke-Sync} is
directly typed with the return value $\frt'$ of the method and with
the corresponding behavioural type. The rule enforces that the cog of
the callee coincides with the local one.

\begin{figure}[t]
{\footnotesize
\[
  \begin{array}{c}
    \mathrule{T-Assign}{
      \typeee{\Gamma}{\nt{rhs}}{\frx}{\cntcel \triangleright \Gamma'}
    }{
      \typest{\Gamma}{x = \nt{rhs}}{\cntcel \triangleright 
        \Gamma'[x\mapsto \frx]} 
    }
    \qquad
    \mathrule{T-Job}{
      \typepe{\Gamma}{e}{\se}  
      \quad
      \typepe{\Gamma}{\this}{\cogty{c}{\se'}} 
    }{
      \typest{\Gamma}{\job{e}}{\se/\se' \triangleright 
        \Gamma} 
    }
    \\\\
    \mathrule{T-Return}{
      \typepe{\Gamma}{e}{\frt}
      \quad
      \typepe{\Gamma}{\destiny}{\frt}
    }{
      \typest{\Gamma}{\ereturn e}{\btnull \triangleright 
        \Gamma} 
    }
    \qquad
    \mathrule{T-Seq}{
    	\begin{array}{c}
      \typest{\Gamma}{s}{{\cal C}[\cntcel_1 \triangleright
        \Gamma_1] \cdots [\cntcel_n \triangleright \Gamma_n]} 
      \\
      \typest{\Gamma_i}{s'}{\cntb_i'}
      \end{array}
    }{
      \typest{\Gamma}{s\semi s'}{{\cal C}[\cntcel_1 \fatsemi
        \cntb_1'] \cdots [\cntcel_n \fatsemi \cntb_n'] } 
    }
    \\\\
    \mathrule{T-If-Nse}{
      \typepe{\Gamma}{e}{\unit}
      \quad
      \typest{\Gamma}{s}{\cntb}
      \quad
      \typest{\Gamma}{s'}{\cntb'}
    }{
      \typest{\Gamma}{\eif{e}{s}{s'}}{\cntb + \cntb'} 
    }
    \qquad
    \mathrule{T-If-Se}{
      \typepe{\Gamma}{e}{\se}
      \quad
      \typest{\Gamma}{s}{\cntb}
      \quad
      \typest{\Gamma}{s'}{\cntb'}
    }{
      \typest{\Gamma}{\eif{e}{s}{s'}}{(\se) \{\cntb\} +
        (\neg\se) \{\cntb'\}}  
    }
  \end{array}
  \]}
  \caption{Typing rules for statements}
  \label{fig:typesystem-stm}
\end{figure}
The typing rules for statements are presented in
Figure~\ref{fig:typesystem-stm}.
The behavioural type in rule \rn{T-Job} expresses the time consumption for  an object with 
capacity $\se'$ to perform $\se$ processing cycles: this time is given by 
$\se/\se'$, which we observe is in general
 a rational number. We will return to this point in Section~\ref{sec.analysis}.

The typing rules for method and class declarations are shown 
in Figure~\ref{fig:typesystem-decl}. 
\begin{figure}[t]
{\footnotesize  \[
  \begin{array}{c}
      \mathrule{T-Method}{
      \begin{array}{c}
      \Gamma({\tt m}) = (\frt_t, \vect{\frt}
      ) \rightarrow \frt_r
      \\
      \typest{\Gamma[\this \mapsto \frt_t]
        [\destiny \mapsto \frt_r][\vect{x} \mapsto \vect{\frt}]
        }{s}{{\cal C}[\cntcel_1 \triangleright 
        \Gamma_1] \cdots [\cntcel_n \triangleright \Gamma_n]}
        \end{array}
    }{
      \typepe{\Gamma}{T\  {\tt m}\ (\vect{T\ x })\ 
        \wrap{s}}{\mty{\m}{\frt_t, \vect{\frt}}
         \wrap{{\cal C}[\cntcel_1 \triangleright
          \varnothing] \cdots [\cntcel_n \triangleright \varnothing]} : \frt_r}
    }
    \\
    \\
    \mathrule{T-Class}{
      \typepe{\Gamma}{\vect{M} }{\vect\mcntc} \quad
    \typest{\Gamma[\ethis\mapsto \cogty{\mathrm{start}}{\const}][\bar{x}\mapsto \bar{\frt}]}{s}{{\cal C}[\cntcel_1 \triangleright
        \Gamma_1] \cdots [\cntcel_n \triangleright \Gamma_n]}
    }{
      \typepe{\Gamma}{	\vect{M}\  \wrap{\vect{T\ x\ssemi}\ s} \
        \key{with}\ \const\ }{\vect\mcntc, {\cal C}[\cntcel_1 \triangleright
        \emptyset] \cdots [\cntcel_n \triangleright \emptyset] }
    }
  \end{array}
  \] }
  \caption{Typing rules for declarations}
  \label{fig:typesystem-decl}
\end{figure}
\paragraph{Examples}~\label{sec.fib.typ}
The behavioural type of the \texttt{fib} method discussed in
Section~\ref{sec:semantics} is 
{\scriptsize
\begin{lstlisting}[mathescape]
fib(c[x],n) {
   (n $\leq$ 1){ $\btnull \triangleright \emptyset$ }
   +
   (n $\geq$ 2){
    1/x $\fatsemi$ d[x] $\fatsemi$ $\nu f{:}~$fib(c[x],n-1)$\to\unit\ \fatsemi$ $\nu g{:}~$fib(d[x],n-2)$\to\unit\ \fatsemi$ 
    $\rmel{f}$$\fatsemi$ $\rmel{g}$$\fatsemi$$\btnull\triangleright\emptyset$
 }
} : $\unit$
 \end{lstlisting}}

\section{The time analysis}
\label{sec.analysis}

The behavioural types returned by the system defined in Section~\ref{sec:typesystem} 
are used to compute upper bounds of time complexity of a
{\coreABS} program.
This computation is performed by an off-the-shelf solver -- the {\tt CoFloCo} solver~\cite{FMH14} -- and, in this section,
we discuss the translation of a behavioural type program into a set of
\emph{cost equations} that are fed to the solver. These cost equations are terms 
\begin{center}
  $m(\vect{x}) = \nt{exp} \quad [\se]$
\end{center}
where 
$m$ is a (cost) function symbol, 
$\nt{exp}$ is an
expression that may contain (cost) function symbol applications (we do not 
define the syntax of $\nt{exp}$, which may be derived by the following
equations; the reader may refer to~\cite{FMH14}), 
and $\se$ is a size expression whose variables are
contained in $\vect{x}$.
%
Basically, our translation maps method types into cost equations, where
(i) method invocations are translated into function applications, and
(ii) cost expressions $\se$ occurring in the types are left unmodified.
%
The difficulties of the translation is that the cost equations must account
for the parallelism of processes in different cogs and for sequentiality of
processes in the same cog. For example,
in the following code:
%
%

{\scriptsize
\begin{lstlisting}[mathescape]
      x = new Class with c; y = new Class with d;
      f = x!m(); g = y!n(); u = g.get ; u = f.get ;
\end{lstlisting}}

\noindent
the invocations of {\tt m} and {\tt n} will run in parallel, therefore their cost 
will be ${\tt max}(t,t')$, where 
$t$ is the time of executing \absinline{m} on \absinline{x}  and $t'$ is the time
executing {\tt n} on \absinline{y}.
%
On the contrary, in the code

{\scriptsize
\begin{lstlisting}[mathescape]
      x = new local Class; y = new local Class;
      f = x!m(); g = y!n(); u = g.get; u = f.get;
\end{lstlisting}}

\noindent
the two invocations are queued for being executed on the same
cog. Therefore the time needed for executing them will be $t+t'$,
where $t$ is time needed for executing \absinline{m} on \absinline{x},
and $t'$ is the time needed for executing \absinline{n} on
\absinline{y}.  To abstract away the execution order of the
invocations, the execution time of \emph{all unsynchronized} methods
from the same cog are taken into account when one of these methods is
synchronized with a \name{get}-statement.  To avoid calculating the
execution time of the rest of the unsynchronized methods in the same
cog more than necessary, their estimated cost are ignored when they
are later synchronized.

In this example, when the method invocation \absinline{y!n()} is
synchronized with \texttt{g.get}, the estimated time taken is $t +
t'$, which is the sum of the execution time of the two unsynchronized
invocations, including the time taken for executing \absinline{m} on
\absinline{x} because both \absinline{x} and \absinline{y} are
residing in the same cog.  Later when synchronizing the method
invocation \absinline{x!m()}, the cost is
considered to be \emph{zero} because this invocation has been taken
into account earlier.

%

\paragraph{The translate function.}
The translation of behavioural types into cost equations is carried out by the 
function \texttt{translate}, defined below.  This function parses
atoms, behavioural types or declarations of methods and classes.  We will use
the following auxiliary function that removes cog
names from (tuples of) $\frt$ terms:
\[
\strip{\_} = \_ \qquad \strip{e} = e \qquad \strip{\cogty{c}{e}} = e
\qquad \strip{\frt_1, \ldots , \frt_n} = \strip{\frt_1}, \ldots , \strip{\frt_n}
\]
We will also use \emph{translation environments}, ranged over by $\Psi$, $\Psi'$, $\cdots$,
which map future names to pairs $(e, \m(\vect{\frt}))$ that records the (over-approximation 
of the) time when the method has been invoked and the invocation. 

In the case of atoms, \texttt{translate} takes four inputs: 
a \emph{translation environment}~$\Psi$, the cog name of the carrier, 
an over-approximated cost $e$ of an execution 
branch, and the atom $\cntcel$.
In this case,  \texttt{translate} returns an updated translation environment and the 
cost. It is defined as follows. 
{\footnotesize
\[
\begin{array}{l}
\atranslate(\Psi, c, e, \cntcel) \; = 
\\ 
\qquad \left\{ \begin{array}{l@{\qquad}l}
        (\Psi, e + e') & \text{when } \cntcel=e'
        \\[.4em]
        (\Psi, e) & \text{when } \cntcel= \nucogty{c}{e'}
        \\[.4em]
        (\Psi, e + \mty{\m}{\strip{\vect{\frt}}}) & \text{when }
        	\cntcel=\mty{\m}{\vect \frt} \rightarrow \frt'
        \\[.4em]
        (\Psi[f\mapsto (e, \mty{\m}{\vect{\frt}})], \, e)
        	& \text{when }  \cntcel=(\newel{f}{\mty{\m}{\vect \frt} \rightarrow \frt'})
       \\[.4em]
       (\Psi\setminus F, e+e_1)))
        	& \text{when } \cntcel=\rmel{f} \quad \text{and} \quad \Psi(f) = (e_f, \mty{\m_f}{c[e'], \vect{\frt_f}})
        \\
         & \text{let } F = \wrap{g \mid \Psi(g) = (e_g, \, \mty{\m_g}{c[e'], \vect{\frt_g}}) }
         \text{ then}
          \\
         & \text{and } e_1 = \sum \wrap{ \mty{\m_g}{\strip{\vect{\frt_g'}}} \; | \; 
           (e_g, \mty{\m_g}{\vect{\frt_g'}}) 
          \in \Psi(F) }
	\\[.4em]
       (\Psi\setminus F, \textit{max}(e, e_1+e_2))
        	& \text{when } \cntcel=\rmel{f} \; \text{and} \; \Psi(f) = (e_f, \mty{\m_f}{c'[e'], \vect{\frt_f}}) \; \text{and} \;  c\neq c'
        \\
         & \text{let } F = \wrap{g \mid \Psi(g) = (e_g, \, \mty{\m_g}{c'[e'], \vect{\frt_g}}) }
         \text{ then}
          \\
          &
          e_1 = \sum \wrap{ \mty{\m_g}{\strip{\vect{\frt_g'}}} \; | \; (e_g, \mty{\m_g}{\vect{\frt_g'}}) 
          \in \Psi(F) } 
          \\
          & \text{and } e_2 = {\it max}\wrap{ e_g \; | \; (e_g, \mty{\m_g}{\vect{\frt_g'}}) 
          \in \Psi(F) } 
          \\[.4em]
       (\Psi, e) &  \text{when } \cntcel=\rmel{f} \; \text{and} \; f \notin \dom(\Psi)
      \end{array} \right.
\end{array}
\]}
%
\noindent
The interesting case of \texttt{translate} is when the atom is $\rmel{f}$. There are three
cases: 
\begin{enumerate}
\item \label{translate.atom.case.samecog} The synchronization is with
  a method whose callee is an object of the same cog. In this case its
  cost must be \emph{added}. However, it is not possible to know when
  the method will be actually scheduled. Therefore, we sum the costs
  of all the methods running on the same cog (worst case) -- the set
  $F$ in the formula -- and we remove them from the translation
  environment.
\item \label{translate.atom.case.diffcog} The synchronization is with
  a method whose callee is an object on a different cog $c'$.  In this
  case we use the cost that we stored in $\Psi(f)$. Let $\Psi(f) =
  (e_f, \mty{\m_f}{c'[e'], \vect{\frt_f}})$, then $e_f$ represents the
  time of the invocation. The cost of the invocation is therefore $e_f
  + \mty{\m_f}{e', \strip{\vect{\frt_f}}}$. Since the invocation is
  \emph{in parallel} with the thread of the cog $c$, the overall cost
  will be ${\it max}(e, e_f + \mty{\m_f}{e',
    \strip{\vect{\frt_f}}})$. As in
  case~\ref{translate.atom.case.samecog}, we consider the worst
  scheduler choice on $c'$.  Therefore, instead of taking $e_f +
  \mty{\m_f}{e', \strip{\vect{\frt_f}}}$, we compute the cost of all
  the methods running on $c'$ -- the set $F$ in the formula -- and we
  remove them from the translation environment.
\item \label{translate.atom.case.nofut}
The future does not belong to $\Psi$. That is the cost of the
invocation which has been already
computed. In this case, the value $e$ does not change.
\end{enumerate}


In the case of behavioural types,  \texttt{translate} takes as input a translation environment, 
the cog name of the carrier, an over-approximated cost of the current
execution branch $(e_1)e_2$, where $e_1$ indicates the conditions corresponding to the 
branch, and the behavioural type $\cntcel$.
%
%
%
%
%
%
%
%
%
%
%
%
{\footnotesize
\[
\begin{array}{l}
  \atranslate(\Psi, c, (e_1)e_2, \cntb) \; = 
\\
\qquad \left\{ 
	\begin{array}{l@{\qquad}l}
   \wrap{ (\Psi',(e_1)e_2') } & \text{when}\ \cntb = \cntcel \triangleright \Gamma
    \quad \text{and} \quad \atranslate(\Psi, c, e_2, \cntcel) = (\Psi',e_2')
    \\\\
    C & \text{when}\ \cntb=\cntcel\fatsemi\cntb' \quad\text{and}\quad \atranslate(\Psi,c, e_2, \cntcel)=(\Psi',e_2')
	\\ & \text{and}\ \atranslate(\Psi', c, (e_1)e_2', \cntb')=C
    \\\\
    C \cup C' & \text{when}\ \cntb=\cntb_1+\cntb_2
   	\quad \text{and}\quad \atranslate(\Psi, c, (e_1)e_2, \cntb_1)= C
    \\
    & \text{and}\quad \atranslate(\Psi, c, (e_1)e_2, \cntb_2)= C'
    \\\\
    C & \text{when}\ \cntb=(e) \wrap{\cntb'}
    \quad \text{and}\quad \atranslate(\Psi, c, (e_1\land e)e_2, \cntb')=C
    \end{array} \right.
\end{array}
\]}%
The translation of the behavioural types of a method is given below. Let $\dom(\Psi) =
\wrap{ f_1 , \cdots , f_n}$. Then we define
$\rmel{\Psi} \eqdef \rmel{f_1} \fatsemi  \cdots  \fatsemi \rmel{f_n}$.
{\footnotesize
\[
\begin{array}{l}
\atranslate(\mty{\m}{\cogty{c}{e},\vect{\frt}}\wrap{\cntb} : \frt) \; = 
\quad
\left[ \quad \begin{array}{l}
			\mty{\m}{e, \vect{e}} = e_1' + e''_1\qquad [e_1]
			\\
			\qquad\qquad \vdots
			\\
			\mty{\m}{e, \vect{e}} = e_n' + e''_n\qquad [e_n]
			\end{array} 
\right.
\\
\\

  \begin{array}{l}
\text{where }\atranslate(\varnothing, c, \btnull, \cntb) = \wrap{\Psi_i,(e_i) e_i'
      \mid 1 \leq i \leq n}, \text{ and } \vect{e} =
    \strip{\vect{\frt}},\\
    \text{and } e''_i= \atranslate(\Psi_i, c, \btnull, \rmel{\Psi_i} \triangleright \emptyset) \; . 
  \end{array}
\end{array}
\]}

\noindent
In addition, $[e_i]$ are the conditions for branching the possible execution
paths of method $\mathtt{m}(e, \many{e})$, and $e_i'+e''_i$ is the
over-approximation of the cost for each path. In particular, $e_i'$
corresponds to the cost of the synchronized operations in each path
(e.g., $\key{job}$s and $\get$s), while $e''_i$ corresponds to the cost of
the asynchronous method invocations triggered by the method, but not
synchronized within the method body.

\paragraph{Examples}~\label{sec.fib.trans}
We show the translation of the behavioural type of fibonacci
presented in Section~\ref{sec.fib.typ}.
Let $\cntb = (\se) \{\btnull \triangleright \emptyset\} + (\neg\se)\{\cntb'\}$,
  where $\se = (\text{n}\leq1)$ and
    $\cntb' = 1/e \ \fatsemi\ \newel{f}{\mathtt{fib}(\cogty{c}{e},
    n-1)}\to \unit\ \fatsemi\
  \newel{g}{\mathtt{fib}(\cogty{c'}{e},n-2)}\to \unit\
    \fatsemi\ \rmel{f}\ \fatsemi\ \rmel{g}\ \fatsemi\ \btnull\
    \triangleright\ \emptyset\}$.
Let also $\Psi=\Psi_1\cup\Psi_2$, where
$\Psi_1=[f \mapsto (1/e, \mathtt{fib}(e, n-1))]$ and
$\Psi_2= [g \mapsto (1/e, \mathtt{fib}(e, n-2))]$.

The following equations summarize the translation of the behavioural
type of the fibonacci method.

{\footnotesize \[
\begin{array}{l}
\atranslate(\emptyset, c, 0, \cntb) 
\\
\quad = \;
\atranslate(\emptyset, c, 0, (\se)\ \wrap{\btnull \triangleright
         \emptyset})\ \cup\
\atranslate(\emptyset, c, 0, (\neg\se)\ \wrap{\cntb'}) 
\\
\quad = \; \atranslate(\emptyset, c, (\se)0, \wrap{\btnull \triangleright \emptyset})\ \cup \ 
\atranslate(\emptyset, c, (\neg\se)0, \wrap{1/e\ \fatsemi\
     \ldots}) 
\\
\quad = \;  \set{(\se)0}\ \cup \ \atranslate(\emptyset, c, (\neg\se)(1/e),
                       \wrap{\newel{f}{\mathtt{fib}(\cogty{c}{e},
                           n-1)}\to\unit\ \fatsemi\ldots})
\\
\quad = \;  \set{(\se)0}\ \cup  \    \atranslate(\Psi_1, c, (\neg\se)(1/e),
                            \wrap{\newel{g}{\mathtt{fib}(\cogty{c'}{e},
                                n-2)}\to\unit\ \fatsemi\ldots})
\\
\quad = \; \set{(\se)0}\ \cup \ \atranslate(\Psi, c, (\neg\se)(1/e),
                    \wrap{\rmel{f}\fatsemi\rmel{g}\fatsemi\ldots})
\\
\quad = \; \set{(\se)0}\ \cup \ \atranslate(\Psi_2, c, (\neg\se)(1/e+ \mathtt{fib}(e, n-1)),
           \wrap{\rmel{g}\fatsemi\ldots})
\\
\quad = \; \set{(\se)0}\ \cup \ \atranslate(\emptyset, c, (\neg\se)(1/e+\max(\mathtt{fib}(e, n-1),\mathtt{fib}(e, n-2))), \wrap{\btnull \triangleright
     \emptyset})
\\
\quad = \; \set{(\se)0}\ \cup \ \wrap{(\neg\se)(1/e+\max(\mathtt{fib}(e, n-1),\mathtt{fib}(e, n-2)))}
\end{array}
\]
     
\[
\begin{array}{l}
\atranslate(\emptyset, c, 0, 0) \; = \; (\emptyset, 0)
\\ 
\atranslate(\emptyset, c, 0, 1/e) \; = \; (\emptyset, 1/e) 
\\
\atranslate(\emptyset, c, 1/e, \newel{f}{\mathtt{fib}(\cogty{c}{e},
  n-1)\to\unit\ } )
\; =  \; (\Psi_1, 1/e) 
\\
\atranslate(\Psi_1, c, 1/e, \newel{g}{\mathtt{fib}(\cogty{c'}{e},
  n-2)}\to\unit\ )
\; = \; (\Psi, 1/e) 
\\
\atranslate(\Psi, c, 1/e, \rmel{f} )
\; = \; (\Psi_2, 1/e+ \mathtt{fib}(e, n-1)) 
\\
\atranslate(\Psi_2, c, 1/e+ \mathtt{fib}(e, n-1), \rmel{g} )
 \;  = \; (\emptyset, 1/e+\max(\mathtt{fib}(e, n-1),\mathtt{fib}(e, n-2))) 
    \end{array}
\]}

  \begin{rules}[mode=simple]
    \begin{array}{ll}
      \atranslate(\texttt{fib}\ (\cogty{c}{e},n)\wrap{ \cntb } : \unit) \; = 
      \\
      \hspace{.3em}\begin{cases}
        \texttt{fib}(e, n) = 0
        & [\textit{n}\ \leq 1]
        \\[.5em]
        \texttt{fib}(e, n) = 1/e+\max(\mathtt{fib}(e, n-1),\mathtt{fib}(e, n-2))
        & [\textit{n}\ \geq 2]
      \end{cases}
    \end{array}
  \end{rules}

\begin{remark}\label{remarkRational}
Rational numbers are produced by the rule \rn{T-Job} of our type system.
In particular behavioural types may manifest terms $\se/\se'$ where 
$\se$ gives the processing cycles defined by
the \texttt{job} operation  and $\se'$ specifies the number of processing cycles per
unit of time the corresponding cog is able to handle. 
%
Unfortunately, our backend solver -- {\tt CoFloCo} -- cannot  handle
rationals $\se/\se'$ where $\se'$ is a variable. 
This is the case, for instance, of our fibonacci example, where the cost of each
iteration is {\tt 1/x}, where {\tt x} is a parameter. In order to analyse this
example, we need to determine \emph{a priori} the capacity to be a constant -- say {\tt 2} --,
obtaining the following input for the solver:
\begin{verbatim}
eq(f(E,N),0,[],[-N>=1,2*E=1]).
eq(f(E,N),nat(E),[f(E,N-1)],[N>=2,2*E=1]).
eq(f(E,N),nat(E),[f(E,N-2)],[N>=2,2*E=1]).
\end{verbatim}
Then the solver gives the following upper bound:
\begin{verbatim}
nat(N-1)* (1/2).
\end{verbatim}

It is worth to notice that fixing the fibonacci method is easy because the capacity does 
not  change during the evaluation of the method. This is not always the case, as in the 
following alternative definition of fibonacci:

{\scriptsize
\begin{lstlisting}[mathescape]
    Int fib_alt(Int n) {
        if (n<=1) { return 1; }
        else { Fut<Int> f; Class z; Int m1; Int m2;
               job(1);
               z = new Class with (this.capacity*2) ;
               f = this!fib_alt(n-1); g = z!fib_alt(n-2);
               m1 = f.get; m2 = g.get;
               return m1+m2; }
    }
\end{lstlisting}}

\noindent
In this case, the recursive invocation {\tt z!fib\_alt(n-2)} is performed on 
a cog  with twice the capacity of the current one and {\tt CoFloCo} is not 
able to handle it.
%
%
It is worth to observe that this is a problem of the solver, which
is otherwise very powerful for most of the examples.
Our behavioural types carry enough information for dealing with more complex examples, 
so we  will consider alternative solvers or combination of them  for
dealing with examples like {\tt fib\_alt}.
  \end{remark}

\section{Properties}
\label{sec:proofoutline}

In order to prove the correctness of our system, we need 
 to show that
$(i)$  the behavioural type system is correct, and
 $(ii)$ the computation time returned by the solver
   is an upper bound of the actual cost of the computation.

The correctness of the type system in Section~\ref{sec:typesystem} is 
demonstrated by means of a subject reduction theorem expressing that if a
runtime configuration $\nt{cn}$ is well typed and $\nt{cn} \to \nt{cn}'$ then 
$\nt{cn}'$ is well-typed as well, and the computation time of~$\nt{cn}$ is larger or equal
to that of $\nt{cn}'$. In order to formalize this theorem we extend the
typing to configurations and we also use extended behavioural types $\concntc$ with the 
following syntax
%
  {\footnotesize
  \begin{syntax}
    \simpleentry 
    \concntc [] \quad \cntb 
     \bnfor \cinvoc{f}{c}{\cntb } \bnfor \concntc\parallel\concntc
    \qquad \qquad [runtime behavioural type]
  \end{syntax}
}%
\noindent%
The type $\cinvoc{f}{c}{\cntb }$ expresses the behaviour of an
asynchronous method bound to the future~$f$ and running in the cog
$c$; the type $\concntc\parallel\concntc'$ expresses the parallel
execution of methods in~$\concntc$ and in $\concntc'$.

We then define a relation $\cls_t$ between runtime behavioural types
that relates types. The definition is algebraic, and $\concntc \cls_t
\concntc'$ is intended to mean that the computational time of
$\concntc$ is at least that of $\concntc'+t$ (or conversely the
computational time of $\concntc'$ is at most that of $\concntc-t$).
This is actually the purpose of our theorems.

%
%
%
%
%

\begin{theorem}[Subject Reduction]
\label{thm.subjectreduction}
Let $\nt{cn}$ be a configuration of a {\coreABS} program and
let~$\concntc$ be its behavioural type. If $\nt{cn}$ is not strongly
$t$-stable and $\nt{cn}\to \nt{cn}'$ then there exists~$\concntc'$
typing $\nt{cn}'$ such that $\concntc \cls_0 \concntc'$.  If $\nt{cn}$
is strongly $t$-stable and $\nt{cn}\to \nt{cn}'$ then there
exists~$\concntc'$ typing $\nt{cn}'$ such that $\concntc \cls_t
\concntc'$.
\end{theorem}
The proof of is a standard case analysis on the last reduction rule applied.

The second part of the proof requires an extension of the $\atranslate$ function to
runtime behavioural types. We therefore define a cost of the equations ${\cal E}_\concntc$ returned by
 $\atranslate(\concntc)$ -- noted $\fcost({\cal E}_\concntc)$ -- by unfolding the 
equational definitions. 
  
\begin{theorem}[Correctness]
\label{thm.correctness}
If $\concntc \cls_t \concntc'$, then $\fcost({\cal E}_\concntc) \; \geq \; \fcost({\cal E}_{\concntc'})+t$.
\end{theorem}
As a byproduct of Theorems~\ref{thm.subjectreduction} and~\ref{thm.correctness}, we 
obtain the correctness of our technique, modulo the correctness of the solver.

\section{Related work}
\label{sec:relatedwork}

In contrast to the static time analysis for sequential executions
  proposed in~\cite{JohnsenPSTC15}, the paper proposes an approach to
  analyse time complexity for concurrent programs.  Instead of using a
  Hoare-style proof system to reason about end-user deadlines, we
  estimate the execution time of a concurrent program by deriving the
  time-consuming behaviour with a type-and-effect system.

Static time analysis approaches for concurrent programs can be divided
into two main categories: those based on type-and-effect systems and those based on
abstract interpretation -- see references in~\cite{TrinderCHLM13}.
Type-and-effect systems (i) collect constraints
on type and resource variables and (ii) solve these constraints.
The difference with respect to our approach is that we do not perform
the analysis during the type inference. We use the type system for
deriving behavioural types of methods and, in a second phase,
we use them to run a (non compositional) analysis
that returns cost upper bounds. This dichotomy allows us
to be more precise, avoiding  unification of variables that are
performed during the type derivation.
In addition, we notice that the techniques in the literature are devised for programs 
where parallel
 modules of sequential code are running. The concurrency is not part
 of the language, but used for parallelising the execution.

 Abstract interpretation techniques have been proposed addressing
 domains carrying quantitative information, such as resource
 consumption. One of the main advantages of abstract interpretation is
 the fact that many practically useful optimization techniques have
 been developed for it. Consequently, several well-developed automatic
 solvers for cost analysis already exist.  These techniques either use
 finite domains or use expedients (widening or narrowing functions) to
 guarantee the termination of the fix-point generation. For this
 reason, solvers often return inaccurate answers when fed with systems
 that are finite but not statically bounded.
For instance, an abstract interpretation technique  that is very close to our contribution
is~\cite{AlbertCJR15}.
The analysis of this paper targets a language with the same concurrency model as ours, and
the backend solver for our analysis, \texttt{CoFloCo}, is a slightly modified version of the solver used by~\cite{AlbertCJR15}.
However the two techniques differ profoundly in the resulting cost
equations and in the way they are produced.
%
Our technique computes the cost by means of a type system, therefore every method has
an associated type, which is parametric with respect to the arguments. Then these types
are translated into a bunch of cost equations that may be \emph{composed} with those of other
methods. So our approach supports a technique similar to \emph{separate compilation}, and
is able to deal with systems that create statically an unbounded but finite number of nodes. 
On the contrary, the technique in~\cite{AlbertCJR15} is not compositional because it takes 
the whole program and computes the parts that may run in parallel. Then the cost equations
are generated accordingly. This has the advantage that their technique does not have any
restriction on invocations on arguments of methods that are
(currently) present in our one.

We finally observe that our behavioural types may play a relevant role
in a cloud computing setting because they may be considered as
abstract descriptions of a method suited for SLA compliance.


\section{Conclusions}
\label{sec:conclusions}


This article presents a technique for computing the time of concurrent object-oriented
programs by using behavioural types. 
The programming language we have studied features an
explicit cost annotation operation that define the number of machine cycles required
before executing the continuation. The actual computation activities of the
program are abstracted by $\key{job}$-statements, which are the unique
operations that consume time.  The
computational cost is then measured by introducing the notion of
(strong) \mbox{$t$-\emph{stability}} (\emph{cf.}~Definition~\ref{def:stability}),
which represents the ticking of time and expresses that up to $t$ time
steps no control activities are possible.
A Subject Reduction theorem (Theorem~\ref{thm.subjectreduction}),
then, relates this stability property to the derived types by stating
that the consumption of $t$ time steps by \key{job} statements is
properly reflected in the type system. Finally,
Theorem~\ref{thm.correctness} states that the solution of the cost
equations obtained by translation of the types provides an upper bound
of the execution times provided by the type system and thus, by
Theorem~\ref{thm.subjectreduction}, of the actual computational cost.

Our behavioural types are translated into so-called cost equations
that are fed to a solver that is already available in the literature
-- the {\tt CoFloCo} solver~\cite{FMH14}. As discussed in
Remark~\ref{remarkRational}, {\tt CoFloCo} cannot handle rational
numbers with variables at the denominator. In our system, this happens
very often. In fact, the number~$\nt{pc}$ of processing cycles needed
for the computation of a $\job{\nt{pc}}$ is divided by the speed
$\nt{s}$ of the machine running it. This gives the cost in terms of
time of the $\job{\nt{pc}}$ statement. When the capacity is not a
constant, but depends on the value of some parameter and changes over
time, then we get the untreatable rational expression.  It is worth to
observe that this is a problem of the solver (otherwise very powerful
for most of the examples), while our behavioural types carry enough
information for computing the cost also in these cases. We plan to
consider alternative solvers or a combination of them for dealing with
complex examples.

Our current technique does not address the full language. In particular we are still
not able to compute costs of methods that contain invocations to arguments which 
do not live in the same machine (which is formalized by the notion of cog in our language).
In fact, in this case it is not possible to  estimate the cost without any indication 
of the state of the remote machine. A possible solution to this issue is to 
deliver costs of methods that are parametric with respect to the state of remote
machines passed as argument. We will investigate this solution in future work.

%
In this paper, the cost of
a method also includes the cost of
the asynchronous invocations in its body that have not
been synchronized. A more refined analysis, combined with the resource analysis 
of~\cite{garciaLL15},
might consider the cost of each machine, instead of the overall cost. That is,
one should count the cost of a method \emph{per} machine rather than in a
cumulative way. While these values are identical when the invocations are always 
synchronized, this is not the case for unsynchronized invocation and 
a disaggregated analysis might return
better estimations of virtual
machine usage.

 \bibliographystyle{abbrv}
 \bibliography{main}




\end{document}